\documentclass[final,5p,times]{elsarticle}

\usepackage{amsmath,amssymb,amsfonts}
\usepackage{graphicx}
\usepackage{siunitx}

\usepackage{lineno}
\usepackage{xspace}
\usepackage{subcaption}
\newcommand{\tp}{\textsc{TelePix2}\xspace}
\newcommand{\hb}{{fast Hit-OR}\xspace}
\newcommand{\tpone}{\textsc{TelePix}\xspace}
\newcommand{\dtwo}{\textsc{DESY II test beam}\xspace}

\makeatletter
\def\ps@pprintTitle{%
	\let\@oddhead\@empty
	\let\@evenhead\@empty
	\def\@oddfoot{}%
	\let\@evenfoot\@oddfoot}

\makeatother
\journal{Nuclear Instruments and Methods A}

\begin{document}
	\begin{frontmatter}
\title{\tp: Full scale fast region of interest trigger and timing for the EUDET-style telescopes at the DESY II Test Beam Facility }

\author[a]{L.~Huth\corref{cor1}}
		\ead{lennart.huth@desy.de}
\author[b]{H.~Augustin}
\author[b]{L.~Dittmann}
\author[b]{S.~Dittmeier}
\author[c]{J.~Hammerich}
\author[a]{Y.~He}
\author[a]{A.~Herkert}
\author[a]{D.~Kim}
\author[d]{U.~Kr\"amer}
\author[b]{D.~M.~Immig}
\author[b]{R.~Kolb}
\author[e]{I.~Peri\'c}
\author[f]{B.~Pilsl}
\author[g]{T.~Senger}
\author[a]{S.~Ruiz~Daza}
\author[b]{A.~Sch\"oning}
\author[a]{M.~Stanitzki}
\author[b]{B.~Weinl\"ader}
\author[a]{A.~Wintle}

\address[a]{Deutsches Elektronen-Synchrotron DESY, Notkestr. 85, 22607 Hamburg, Germany}
\address[b]{Physikalisches Institut der Universität Heidelberg, INF 226, 69120 Heidelberg, Germany}
\address[c]{University of Liverpool, Liverpool L69 72e, United Kingdom}
\address[d]{Nikhef, Science Park 105, 1098 XG Amsterdam}
\address[e]{Institut für Prozessdatenverarbeitung und Elektronik, KIT, Hermann-von-Helmholtz-Platz 1, 76344 Eggenstein-Leopoldshafen, Germany}
\address[f]{HEPHY - Institut für Hochenergiephysik der Österreichischen Akademie der Wissenschaften, Nikolsdorfer Gasse 18, 1050 Wien, Austria}
\address[g]{Physik-Institut, Universit\"at Z\"urich, Z\"urich, Switzerland}

\begin{abstract}
With increasing demands by future and current upgrades of particle physics experiments on rate capabilities and time resolution, the requirements on test beams are also increasing. 
The current infrastructure at the DESY II test beam facility includes particle tracking telescopes with long integration times, no additional timing but excellent spatial resolution. 
This results in readouts with multiple particles per trigger, causing ambiguities in tracking and assigning particles to triggers. 
Also, it is likely not to trigger on particles that pass through a small device under test, leading to inefficient data taking.
These issues can be solved by adding \tp as a timing and flexible region of interest trigger layer. 
\tp is a full scale HV-CMOS chip based on the successful small scale prototype \tpone.
The DAQ system and the sensors performance featuring efficiencies above \SI{99}{\percent} and a time resolution of  \SI{3.844\pm0.002}{ns} are presented. 
The integration into EUDAQ2 and the AIDA-TLU to seamlessly work in the test beam environment as well as into the analysis chain is described. 
First successful use cases are highlighted to conclude that \tp is a well-suited timing and trigger layer for test beams. 
\end{abstract}
\end{frontmatter}

\section{Introduction}
\label{sec:intro}
Developing new detectors for particle physics experiments is crucial improving the understanding of fundamental forces and interactions. 
Characterisation of novel detectors under conditions as close as possible to the final experiment can be done at test beams, where minimum ionizing particles are provided for tests. 
Precise reference systems - so-called beam telescopes - are used to tag the spatial and temporal presence of a particle. 
DESY hosts a test beam facility~\cite{DIENER2019265} at the DESY II accelerator with three independent beamlines of which two are equipped with EUDET-style beam telescopes~\cite{jansen2016}, based on MINMOSA-26 sensors~\cite{huguo2010,baudot2009} with an active area of approximately \SI[parse-numbers=false]{2x1}{cm\squared}. Copies of these telescopes can also be found at CERN.
These telescopes have an integration time of up to \SI{230}{\micro\second}, no time-stamping and are read out in externally triggered rolling shutter mode. 
The telescope is fully integrated with EUDAQ2~\cite{Liu:2019wim} and the AIDA-TLU~\cite{Baesso_2019}, allowing users to integrate devices under test (DUT) with minimal effort.  \\
Under typical operation conditions, up to 6 electrons pass the telescope in a single readout frame.
This causes ambiguities in track reconstruction. In addition, it is likely not to trigger the particle passing through potentially small DUTs, which results in inefficient data taking.
A pixelated timing layer with a configurable trigger region resolves both issues by assigning a timestamp to each particle passing the telescope and only triggering the telescope readout if the selected region is hit. Previously used hybrid approaches~\cite{Obermann_2014} have reduced timing capabilities, are operated in \SI{40}{MHz} LHC-like readouts and add significant material.
The time resolution of a suitable layer has to be below \SI{1}{\micro\second} to be able to discriminate particles coming from different circulations of the primary beam and below \SI{5}{ns} to provide insight into the timing for LHC like detectors. The active area should correspond to the telescope's active area.
\tp is the successor of a small-scale prototype \tpone \cite{AUGUSTIN2023167947}, that already fulfilled all requirements except for the size.
\section{\tp}
\tp is a system-on-chip pixel detector produced in a \SI{180}{nm} HV-CMOS process and based on the developments within the Mu3e and ATLAS context~\cite{PERIC2007876,SCHIMASSEK2021164812}. 
It has an active area of \SI[parse-numbers=false]{2\times1}{cm\squared} to match the active area of the MIMOSA-26 chips. 
The matrix consists of \SI[parse-numbers=false]{120\times400}{pixels} with a pitch of \SI[parse-numbers=false]{165\times25}{\micro m\squared}.
The chip consists of repeated columns, where one column has 400 pixels point-to-point connected to 400 hit digitizers and end-of-column multiplexer. 
Each pixel can store a 20-bit time stamp as well as a 10-bit time-over-threshold.
Hit data is streamed zero suppressed via a differential link to the readout system typically running at  \SI{1.25}{Gbit/s}. 
The chip periphery additionally hosts a readout control unit, a clock generator based on a phase-locked loop, configuration registers, DACs and I\/O pads.\\
\tp is implemented on a p-type Czochralski substrate with resistivity in the range \SI[parse-numbers=false]{200 - 400}{\Omega cm} instead of the standard wafer to enhance the depleted region and hence charge collection.  \\
A pixel diode is formed by a deep n-well implant which is reversely biased towards the substrate. The pixel is designed to have a breakdown voltage of about 120 V. 
The used technology offers 7 metal layers, with a thick top metal layer used for the distribution of power and time-critical signals.
The active pixel matrix extends to three die borders making the chip three-side abuttable. \\
All configuration register cells are triple redundant and followed by majority logic making it tolerant to single-bit flips caused by ionization (single-event upset). The cells have an auto-refresh feature to allow for the automated repair of single-event upsets.
These registers store settings for the DACs that generate bias-, threshold-, baseline- and injection voltages.

\subsection*{Pixel Cell}
\begin{figure}[h]
    \centering
    \includegraphics[width=.95\columnwidth]{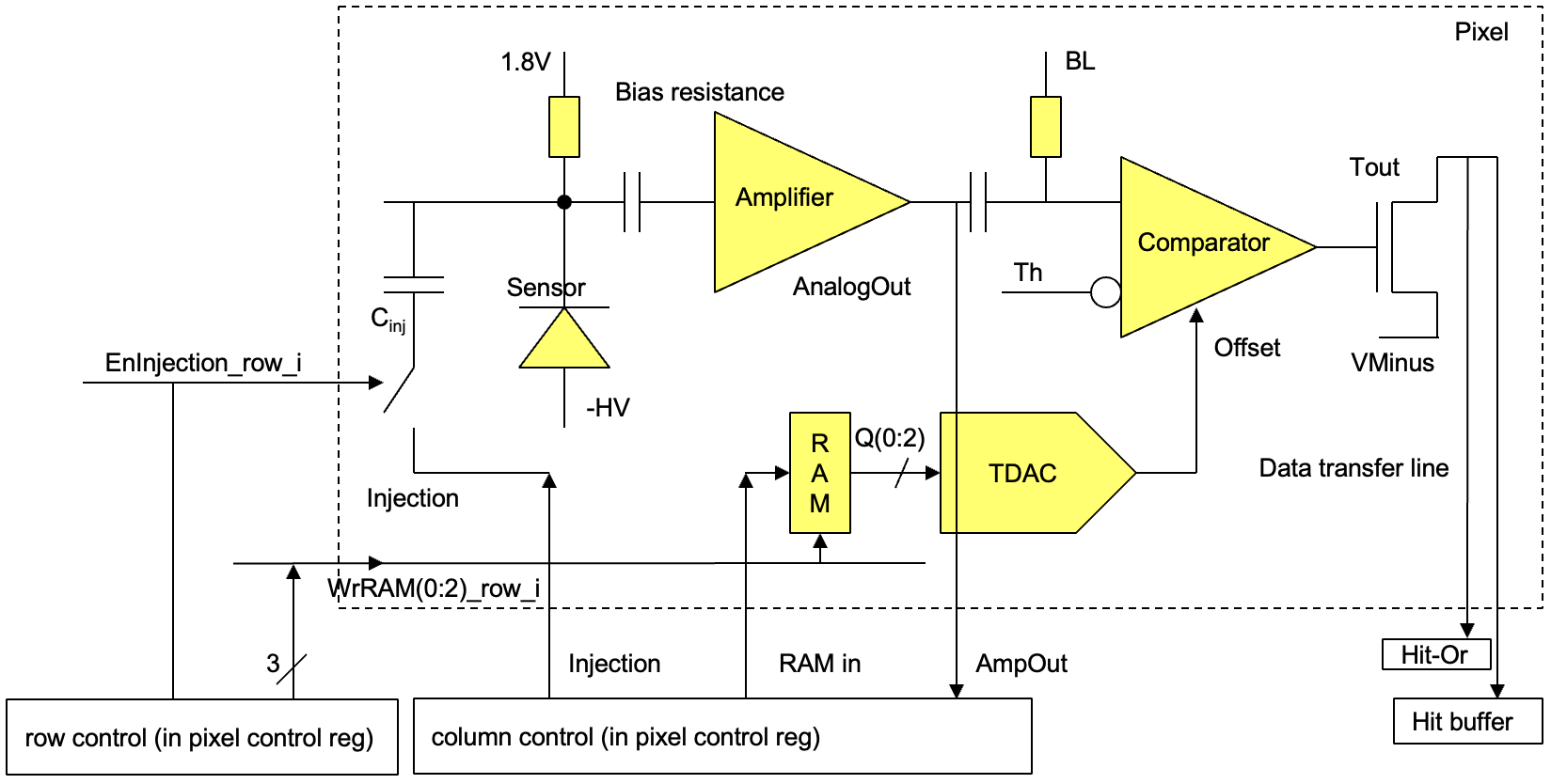}
    \caption{Schematic drawing of the in-pixel functionality}
    \label{fig:pixelCell}
\end{figure}

The main parts of the active pixel cell are the sensor diode - implemented as a large fill factor n-well electrode, the AC-coupled charge sensitive amplifier (CSA) and the comparator. 
The negative signal charge is collected by the n-well and amplified by an inverting CSA. 
No pixel reset is required since the feedback capacitor (p+ diffusion to n-well capacitance of the node AmpOut in Figure~\ref{fig:pixelCell}) of the amplifier is continuously discharged. 
The comparator compares the positive amplifier output signal to a threshold, which is adjustable via a 3-bit differential current-steering DAC, that is stored in an SRAM cell. 
The comparator output signals are transmitted to the hit digitizers at the bottom of the chip using long signal lines. \\
An additional test signal circuit allows the injection of a defined charge into the n-well by discharging a metal-metal capacitor enabling electrical tests without ionizing particles. 
It is produced at the chip periphery, its amplitude can be varied by an on-chip DAC with 8 bits. 
The test signal is distributed to all pixels but can be enabled/disabled per pixel.
The block diagram of the pixel is shown in figure~\ref{fig:pixelCell}.

\subsection*{Hit Digitization}
The in-pixel discriminator signal is transmitted to the hit digitizers, where the rising edge is sampled with a 19-bit Gray counter running at the rising reference clock (typically \SI{125}{MHz}) edge together with a falling clock edge 1-bit counter, allowing for time stamping with twice the clock speed. An additional 10-bit time stamp samples the threshold crossing of the falling signal edge to determine the time-over-threshold. Additionally, the address of the hit is generated. The layout of the hit digitizer is full custom and has an area of \SI[parse-numbers=false]{165x4.2}{\micro\meter\squared}.

\subsection*{Digital Periphery}
The readout process is governed by a synthesized readout control unit (RCU) which comprises configuration registers, the readout state machine, an 8b10b encoder and the first stages of a data serializer realized in a binary tree structure.
The final stage of the serializer is implemented in custom electronics and allows to send out the data at nominally \SI{1.25}{GBit/s}, sending out a bit on the rising and falling edge of the fast base clock.
This full custom part, which also provides clocking, is implemented using differential current mode logic (DCL) instead of CMOS logic.
\tp has a clock generator based on a ring oscillator and a phase-locked
loop (PLL). The PLL receives an externally generated reference clock and generates a 5 times fastefo base clock from which all other required clock signals are derived.
The RCU is verified up to a reference clock speed of 160 MHz.

\subsection{Data acquisition system}
\begin{figure}
   \centering
    \includegraphics[width=\columnwidth]{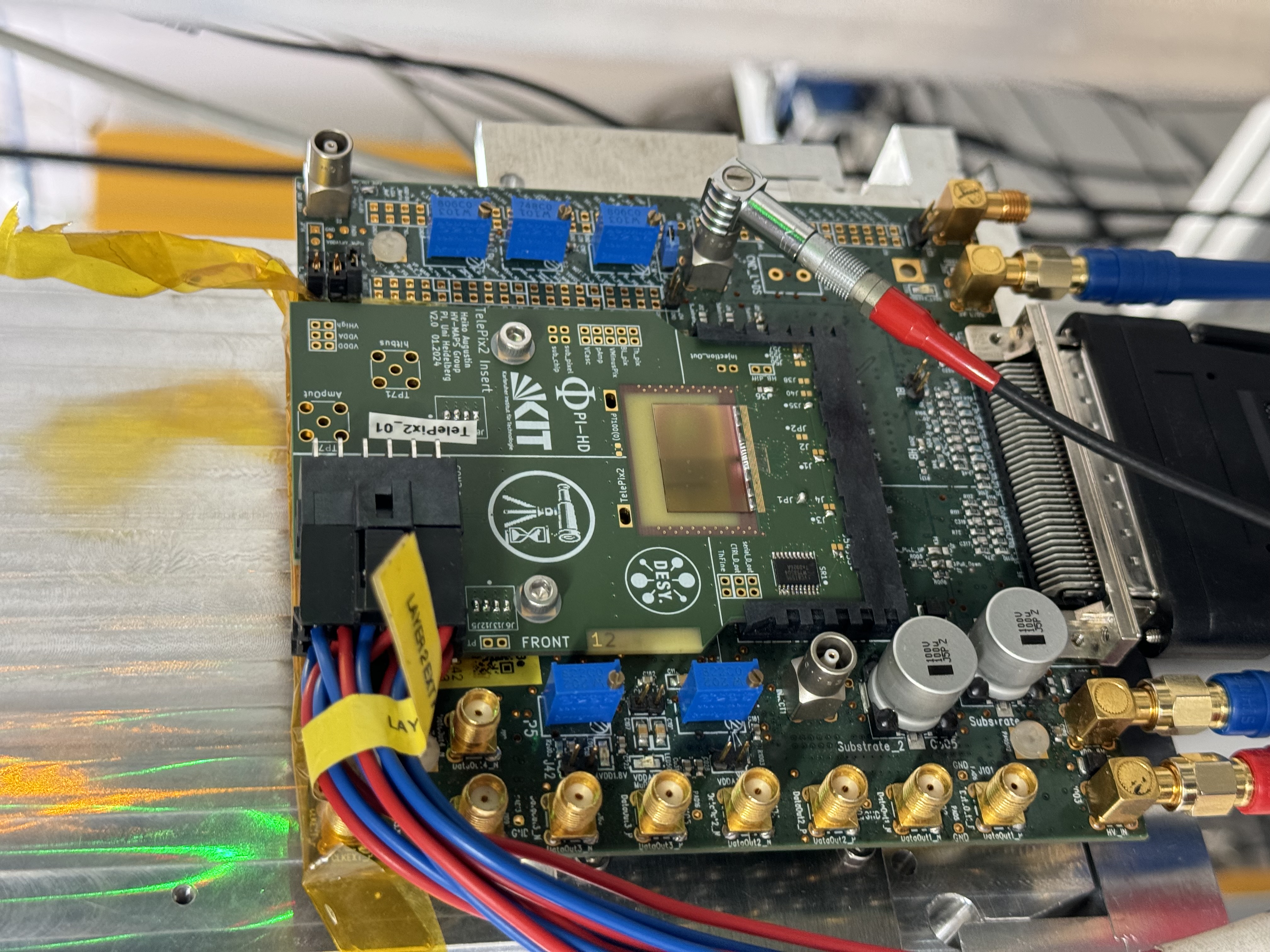}
    \caption{Photograph of the \tp chipboard, mounted to the motherboard. The red Lemo cable connects the fast hit-or, the Molex connector powers the chip, and the SCSI-III cable connects to the readout FPGA.}
    \label{fig:tpBoard}
\end{figure}

The data acquisition system is based on the Heidelberg HV-MAPS laboratory and test beam DAQ system~\cite{TWEPP2017, Dittmeier2018}. 
\tp is glued and wire-bonded to a carrier PCB with power filtering, see figure~\ref{fig:tpBoard}. 
The chipboard connects directly to a power supply providing low voltage to operate the chip with. 
Additionally, it connects to a generic motherboard, which hosts an LVDS repeater to enhance the quality of the differential data stream over a longer cable to the readout FPGA, that is connected via PCIe to the readout computer. 
Data is transferred via direct memory access to the PC's RAM and further processed in the DAQ software: Data is decoded and packed into a frame structure and stored on disk. A fraction of data blocks is forwarded to a monitoring system. 
The DAQ is steered with a Qt-based graphical user interface (GUI), which is described in~\cite{Huth2018}.

\section{Telescope integration}

\subsection*{Data taking setup}
\begin{figure}
    \centering
    \includegraphics[width=\columnwidth]{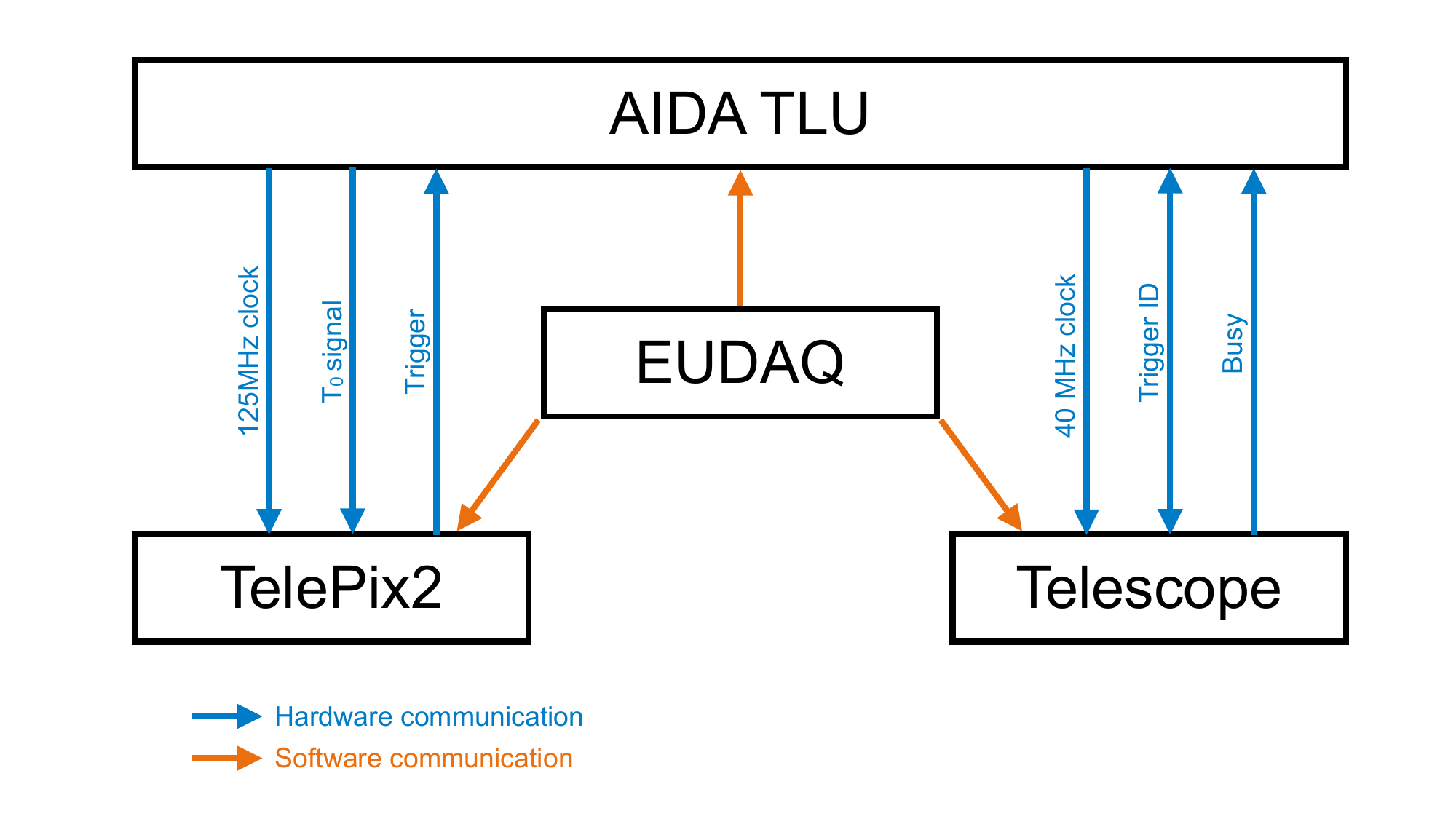}
    \caption{Typical test beam setup. \tp connects to a DUT interface of the AIDA-TLU and receives a 125 MHz clock and T$_0$ at run start while providing a trigger output. The MIMOSA telescope operates in the EUDET mode to exchange the trigger ID, while ADENIUM~\cite{Liu_2023} runs in AIDA+trigger ID mode. EUDAQ2 is used to steer the data taking.
    External DUTs are not shown.}
    \label{fig:setup}
\end{figure}

To integrate \tp with beam telescopes at the facility \dtwo, it must interface with AIDA-TLU and be integrated with the EUDAQ2 framework. 
The latter is achieved via a custom Producer that can be activated via the systems' GUI. Since the data are directly stored on disk, no additional data collectors are required in EUDAQ2. 
Synchronization with the telescope is realised in the synchronous AIDA mode~\cite{Baesso_2019}, where the AIDA-TLU provides a 125 MHz clock to the \tp DAQ, which is forwarded to the chip. 
At run start a T$_0$ signal is sent to \tp which resets all internal counters.
This allows for fully synchronous data-taking without additional interactions between the TLU and \tp. 
The telescopes at DESY exchange a trigger ID with the AIDA-TLU, which stores the timestamp for each trigger. 
The \hb with a rising leading edge, an amplitude of approximately \SI{1.2}{V} and a signal length corresponding to the time-over-threshold can be directly fed into a trigger input of the AIDA-TLU. 
A sketch of a typical test beam setup is shown in figure~\ref{fig:setup}.

\subsection*{Event building with \tp}
\tp can be used as a region of interest (ROI) trigger or timing plane or both simultaneously. Depending on the usage, the event building, illustrated in figure~\ref{fig:events}, works as follows:

\begin{figure}
    \centering
    \includegraphics[width=\columnwidth]{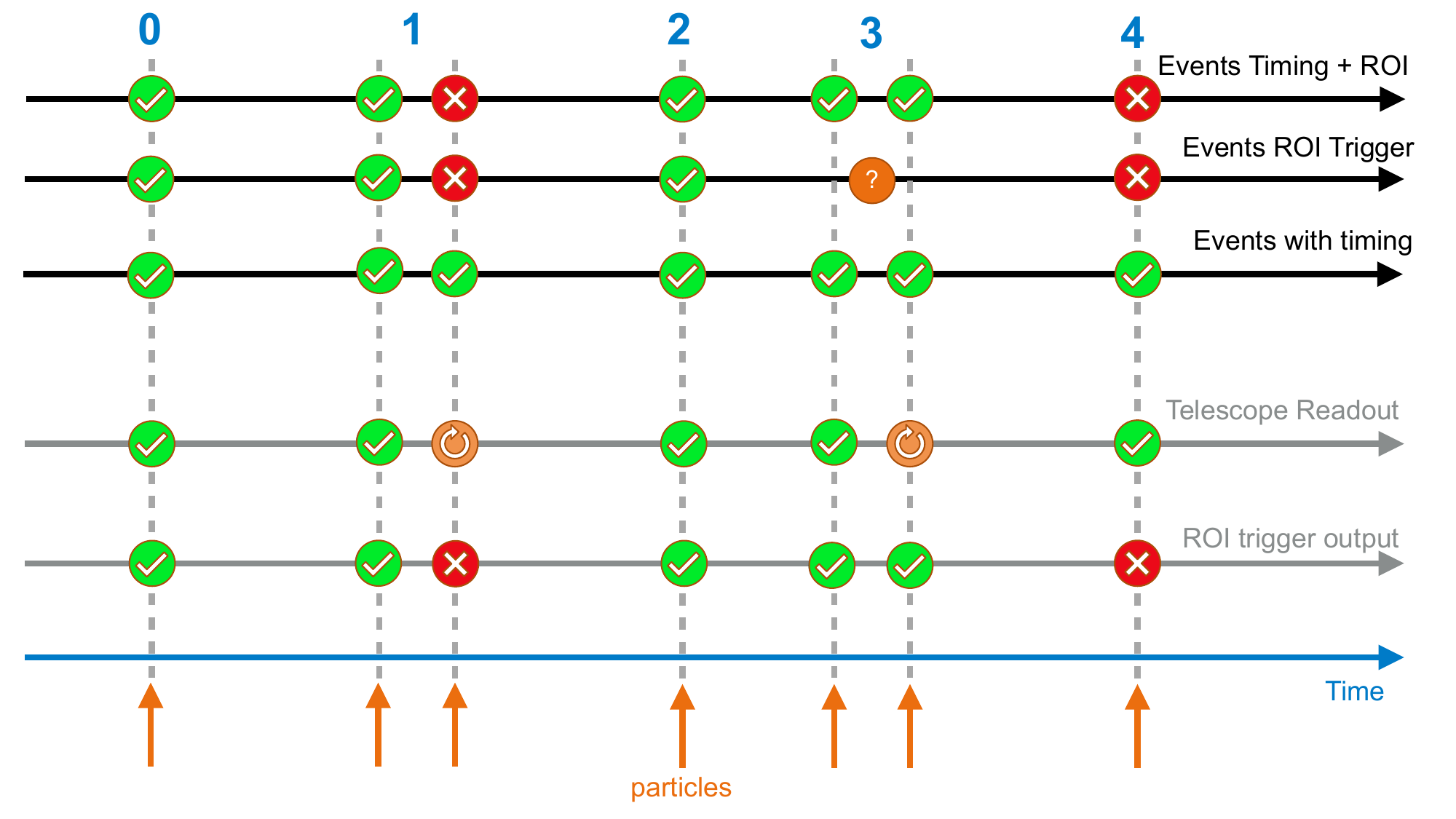}
    \caption{Sketch of the event building with a timing and/or trigger plane. Green check marks indicate correctly assigned particles, red crosses wrongly assigned track time stamps or events that will be rejected, orange question marks ambiguities and the orange loading symbol indicates that a second particle hit is read out in the telescope's frame.}
    \label{fig:events}
\end{figure}

\begin{itemize}
    \item {\bf{Region of interest trigger:}} No additional timestamp information is provided; therefore, only the AIDA-TLU timestamp can be used. In this case, it is essential to record only a single particle per event in the region of interest. An example of a failure of this is shown in event 3 in figure~\ref{fig:events}.
    \item {\bf{Timing plane:}} \tp is used as a timing plane, assigning a time stamp to each hit. This causes efficient suppression of ambiguities but can result in unnecessary triggering, like for event 4 in figure~\ref{fig:events}. 
    \item {\bf{ROI trigger and timing plane:}} In this mode, the full potential of \tp is being exploited. Particles outside the ROI will not trigger a telescope and hence increase data-taking efficiency. If two particles cross the ROI in one event, the time tagging will allow for correct assignment of these tracks.  
\end{itemize}

For user convenience, a custom corryvreckan~\cite{Dannheim_2021} \textsc{EventLoader} is provided to access data for reconstruction. Three exemplary user stories highlighting each of the possible use cases are shown in section~\ref{sec:users}.

\section{Results}
During a test beam campaign at DESY a single \tp has been studied in detail using a \SI{4}{GeV} electron beam as a compromise between particle rate and scattering in the telescope. 
Data recorded is used to study the key performance figures efficiency and time resolution of the full matrix and the timing of the \hb. 
Analysis has been carried out with corryvreckan. 
ADENIUM~\cite{Liu_2023}, an ALPIDE based telescope provided by DESY has been used as a particle tracking reference system.
Since no additional timing plane has been present during data taking with ADENIUM, all \tp hits within the integration time of the reference telescope are accepted for spatial matching. 
A conservative elliptical cut of \SI[parse-numbers=false]{495x75}{\micro\meter\squared} is applied to assign clusters to tracks. 
All presented studies have been performed with a single set of configuration DACs and a supply voltage of \SI{1.8}{V} unless otherwise stated.
The chip has been operated without pixel trimming. 
\subsection*{Hit detection efficiency}
\begin{figure}
         \centering
     \includegraphics[width=\columnwidth]{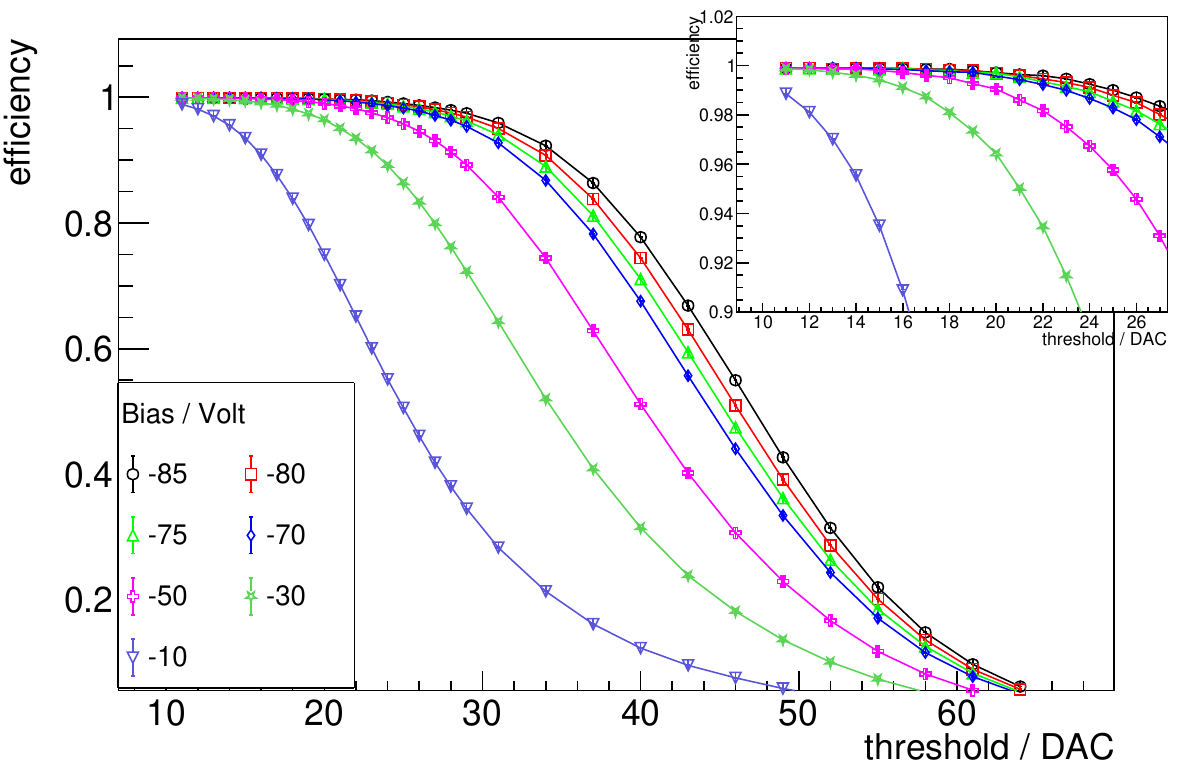}
        \caption{Efficiency as a function of the threshold for different bias voltages. Statistical error bars are too small to be seen.}
    \label{fig:eff_bias}
\end{figure}

The hit detection efficiency is determined as the ratio between reconstructed tracks with an assigned \tp cluster over the total number of tracks intersecting the chip.  
The hit detection efficiency for different bias voltages is summarized in figure~\ref{fig:eff_bias}.

\begin{figure}
         \centering
    \includegraphics[width=\columnwidth]{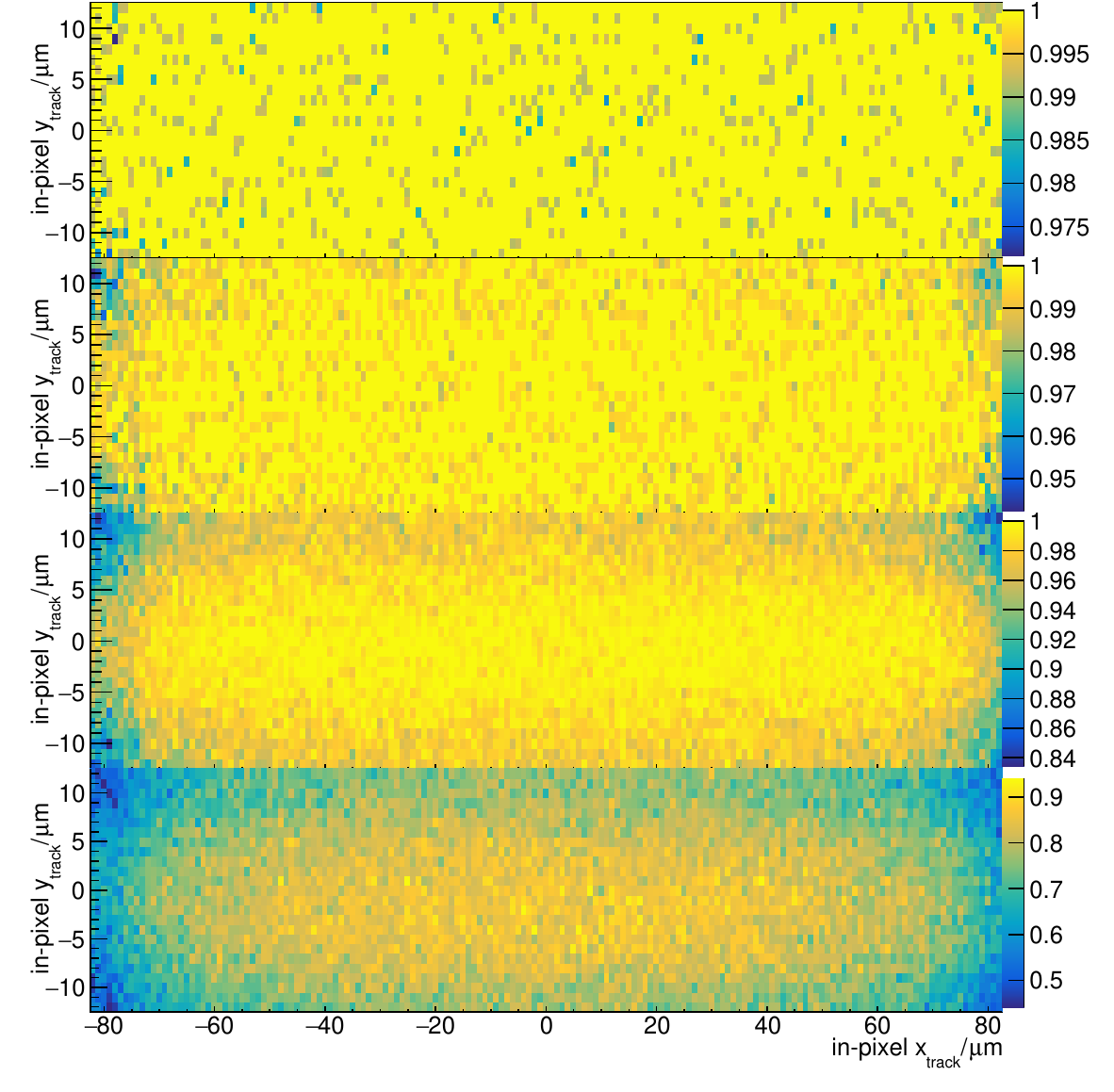}
         \caption{In pixel efficiency of \tp at a bias of \SI{-85}{V}. Thresholds from top to bottom: 16, 21, 28, 40. Please note the change in the z-scale.}
    \label{fig:eff_in_pixel}
\end{figure}
As expected, higher bias increases the threshold range in which the chip can be operated efficiently.
The improvement is less dominant at higher bias voltages as the expected growth in active depleted thickness scales with $\sqrt{HV}$. 
At low thresholds, the hit detection is homogeneous over the pixel area and no position-dependent effects are visible. 
With increasing threshold, the efficiency reduces in the corners first, at the edges second and finally all over the chip as depicted in figure~\ref{fig:eff_in_pixel}.
The initial reduction is explained by charge sharing between up to four pixels in the corners and two at the edges.

\subsection*{Time Resolution}
\begin{figure}
    \centering
    \includegraphics[width=0.8\columnwidth]{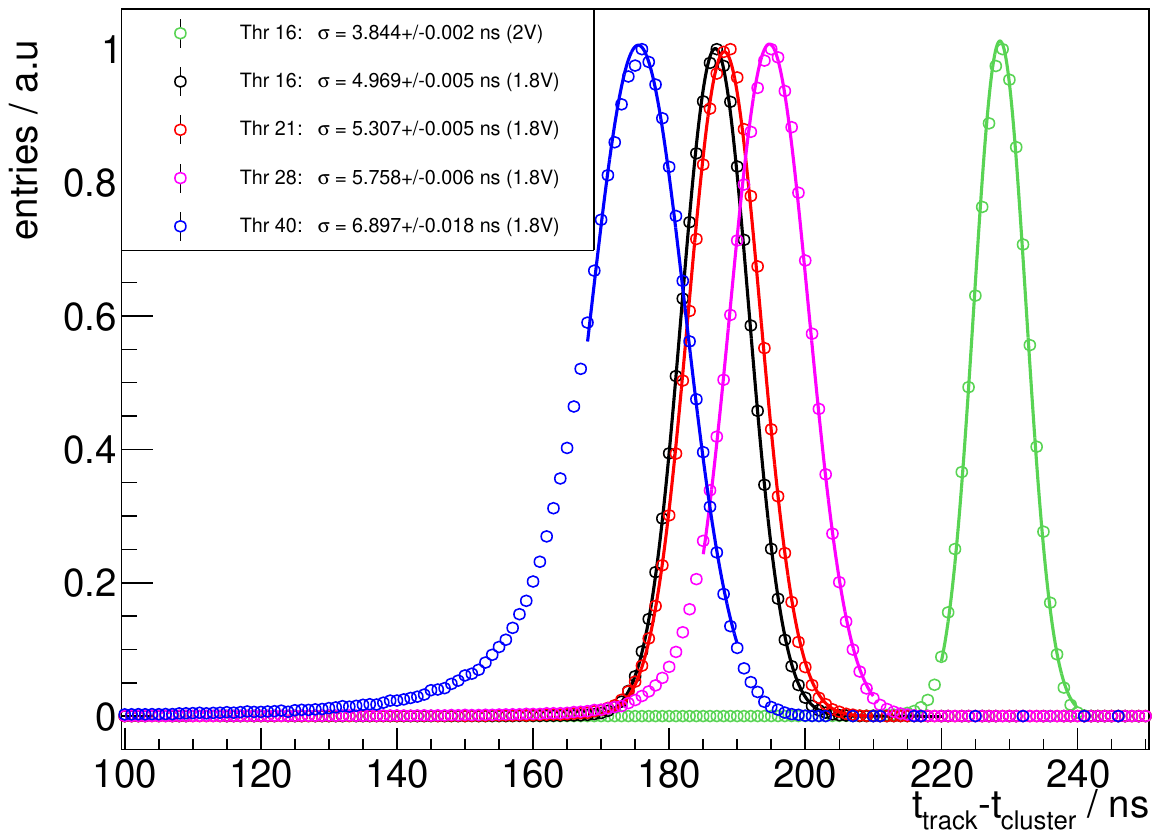}
    \caption{Time difference between reference trigger and \tp cluster time.}
    \label{fig:time-res}
    \end{figure}

\begin{figure}
     \centering
    \includegraphics[width=0.9\columnwidth]{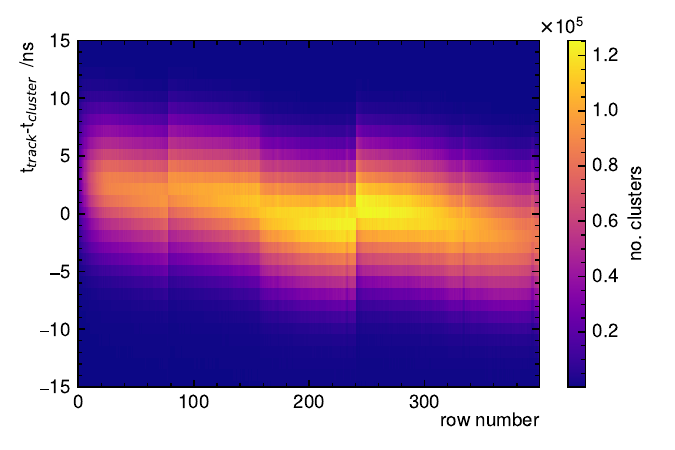}
     \caption{Time difference between reference trigger and \tp cluster time versus the row number.}
    \label{fig:delay}
\end{figure}

The time resolution is determined relative to a trigger scintillator sampled with a \SI{781.25}{ps} timestamp by the AIDA-TLU. 
Only clusters assigned to a track are used, while the cluster time stamp is defined by the pixel timestamp with the largest ToT in a cluster.  
The time resolution of the complete sensor is shown in figure~\ref{fig:time-res} for different detection thresholds.
A Gaussian fit to the core of the distribution leads to a $\sigma$ of \SI{3.844\pm0.002}{ns} at a detection threshold of 16 and a supply voltage of \SI{2}{V}. This is unprecedented for a large area chip in this technology and well within the specifications for a timing layer for the \dtwo.
Data taken at the default \SI{1.8}{V} supply voltage show systematically worse timing performance. 
At threshold 16 a $\sigma$ of  \SI{4.97\pm0.005}{ns} is determined at a larger absolute delay.
With increasing threshold, the time resolution is worsening as expected due to the increased impact of time walk. 
At lower thresholds, better time resolution is expected at the price of increased noise, which is likely not desired if operated as reference sensors and is currently under investigation.
Studying the row dependency of the time residuals, compare figure~\ref{fig:delay}, two effects are observed: Firstly, four sharp cuts are visible. 
These cuts correspond to rows where changes in the metal layers for signal routing occur, which results in different capacitance. 
Secondly, the small slope might be caused by longer transmission lines or voltage drops. 
Correction for these effects and time-walk is a work in progress and will be presented elsewhere. 
    
\subsection*{Region of Interest Triggering}
\begin{figure}
     \begin{subfigure}{\columnwidth}
     \centering
    \includegraphics[width=0.4\columnwidth]{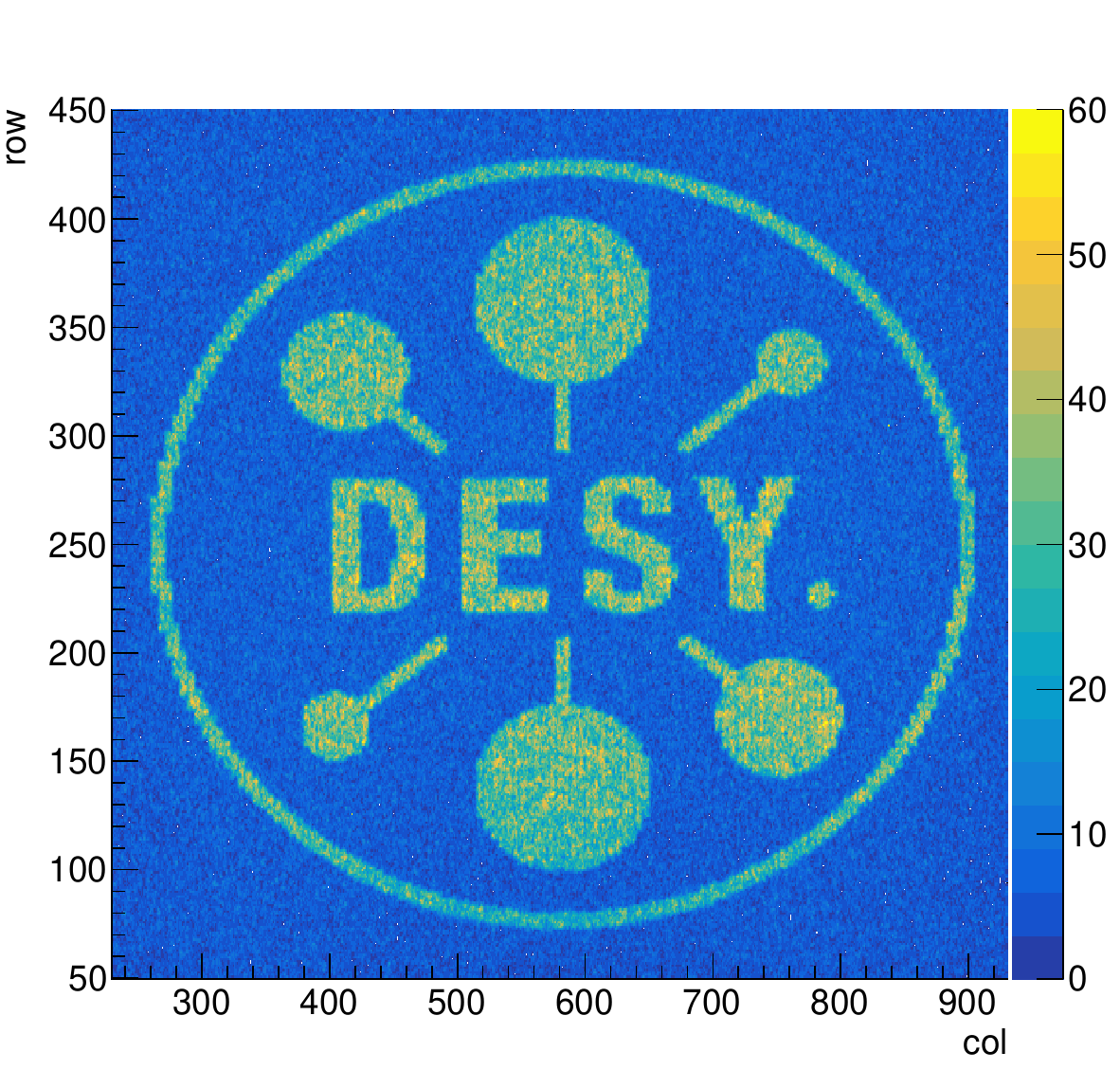}
     \caption{Hitmap of a telescope plane with \tp as trigger. A DESY logo is programmed to the \hb.}
    \label{fig:logo}
     \end{subfigure}
    \begin{subfigure}{\columnwidth}
    \centering
        \includegraphics[width=0.7\columnwidth]{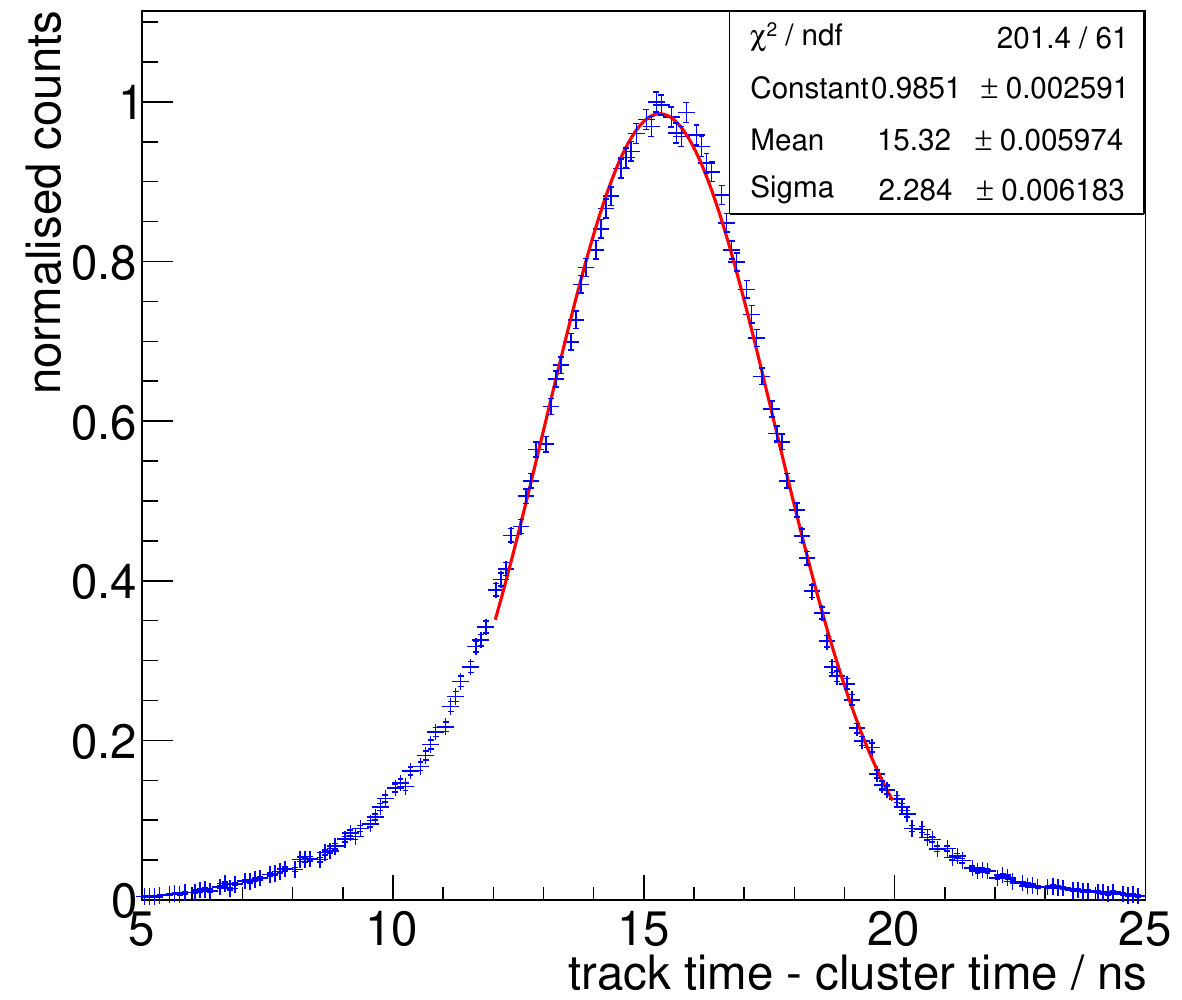}
    \caption{Time difference between reference trigger and \tp hit-or at a threshold of 16.}
    \label{fig:time_res}
    \end{subfigure}
    \caption{The \hb of \tp. }
\end{figure}

The region of interest trigger mode has been evaluated with a supply voltage of \SI{2}{V}. The possibility of selecting an arbitrary pixel arrangement is showcased by only enabling the \hb for pixels assembling a DESY logo and reading out a triggered telescope reference pixel sensor; compare figure~\ref{fig:logo}. Furthermore, the time resolution of the \hb is gauged by comparing it with a scintillator signal coupled to a PMT and read out via an oscilloscope. The result is depicted in figure~\ref{fig:time_res} exemplarily for a quasi noise-free threshold of 16. A Gaussian fit to the distributions' core yields a $\sigma$ of \SI{2.28\pm 0.01}{ns}, sufficient for precise triggering at test beams. The absolute delay of the trigger compared to the scintillator is \SI{15.32\pm 0.01}{ns}, which is an acceptable latency for test beam operation.

\section{User Stories}
\label{sec:users}
\tp is in user operation and utilized by multiple groups to enhance data taking efficiency and suppress ambiguities. In the following three examples using the different operating modes are discussed and the improvement due to \tp is highlighted.

\subsection*{H2M - region of interest trigger mode}
The H2M (Hybrid-to-Monolithic) is implemented in a 65~nm CMOS imaging process as proof of concept for a digital-on-top design workflow.
It ports a hybrid pixel-detector architecture, with digital pulse processing in each pixel, into a monolithic chip. 
The chip matrix consists of 64×16 square pixels, each with a size of \SI{35}{}$\times$\SI{35}{\micro\meter^2}, and a total active area of approximately \SI{1.25}{\meter\meter^2}.
A \SI{100}{MHz} clock is used for non-simultaneous Time-over-Threshold (ToT) or Time-of-Arrival (ToA) measurements.

H2M has been tested in the \dtwo test beam facility using the ADENIUM telescope. \tp is used as region of interest trigger  plane.
A region of interest of approximately \SI{3}{\meter\meter^2} is left unmasked in \tp matching the H2M area with a safety margin. 

\begin{figure}
    \centering
\begin{subfigure}{\columnwidth}
    \includegraphics[width=\columnwidth]{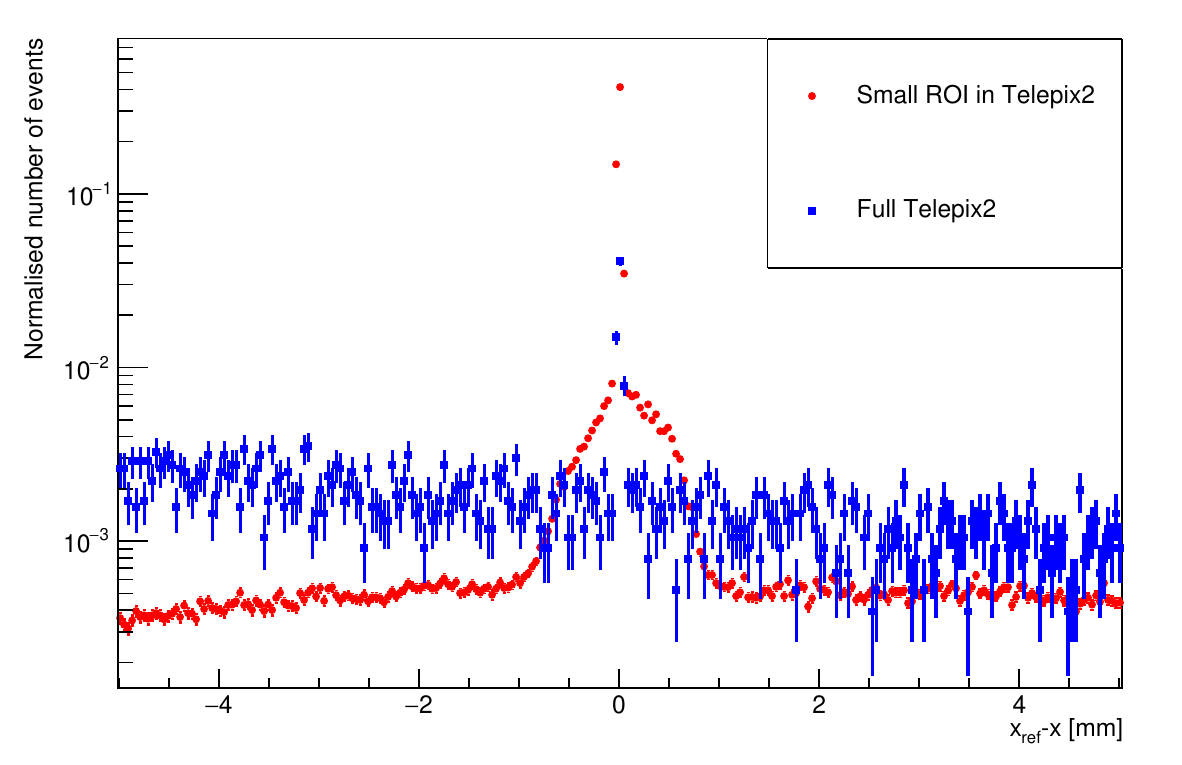}
    \caption{ Spatial correlations of H2M with (red) and without (blue) \tp region of interest trigger along x.}
    \label{fig:correlations_h2m}
    \end{subfigure}
\begin{subfigure}{\columnwidth}
    \includegraphics[width=\columnwidth]{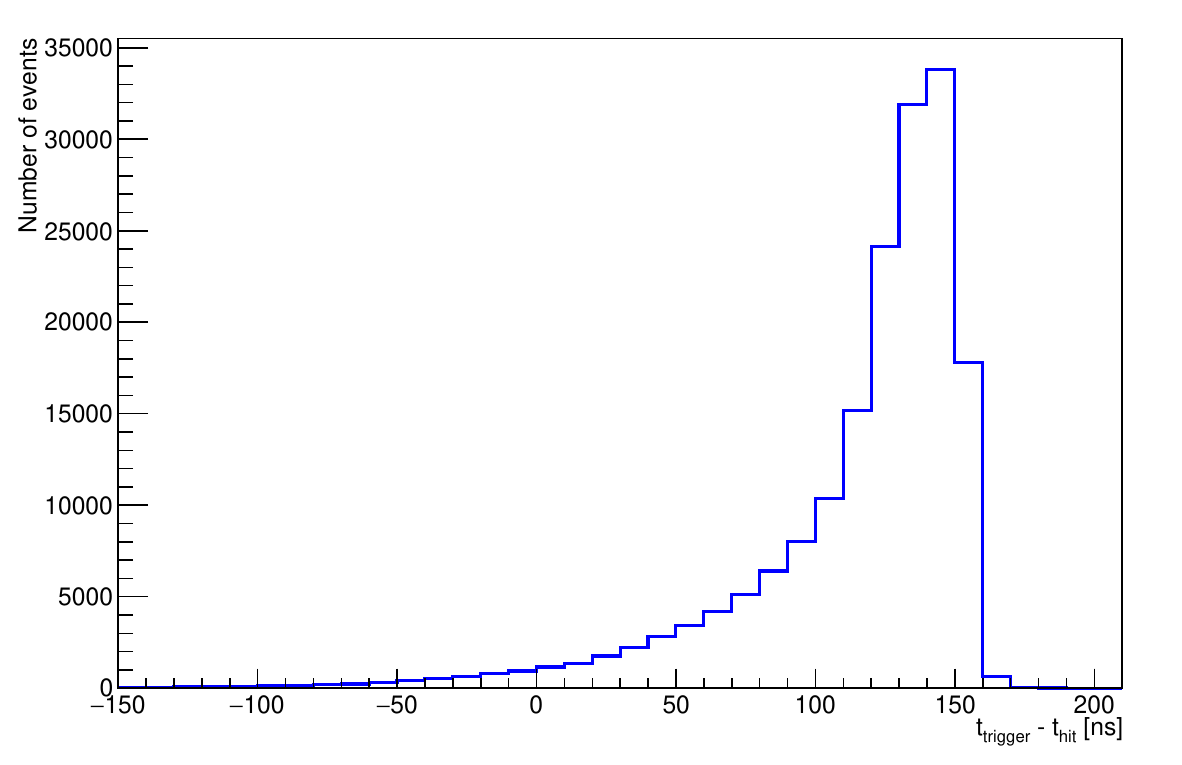}
    \caption{ Temporal correlations between \tp \hb and H2M.}
    \label{fig:timeresidual_h2m}
\end{subfigure}
\caption{Spatial and temporal correlations of H2M \tp.}
\end{figure}

Figure \ref{fig:correlations_h2m} compares the correlations in X between H2M and the reference plane with data taken from a small region of interest and the entire \tp. While a one order of magnitude higher background of uncorrelated events is observed for the full region, for the small region, most of the events are correlated. The latter reduces the number of empty readout frames of H2M and highly improves the data taking efficiency.
Moreover, time studies have been performed based on \tp. Figure~\ref{fig:timeresidual_h2m} shows the time residuals between the \tp\,\hb and H2M time stamps. The tail to the left is caused by time-walk of H2M.

\subsection*{ATLAS ITk strips - timing plane mode}
\begin{figure}
\centering
	\begin{subfigure}{\columnwidth}
		\includegraphics[width=\columnwidth]{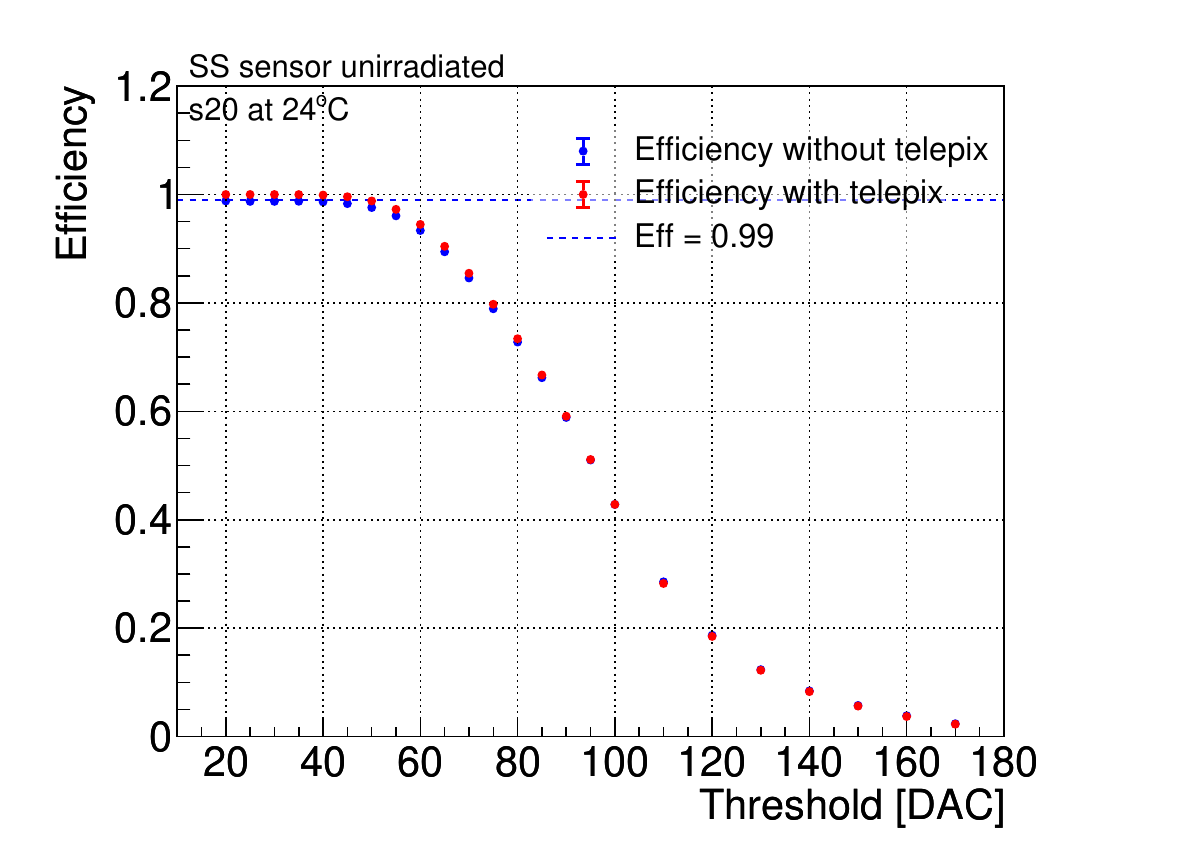}
        \caption{Measured module efficiency versus threshold.}
	\end{subfigure}	
	\begin{subfigure}{\columnwidth}
		\includegraphics[width=\columnwidth]{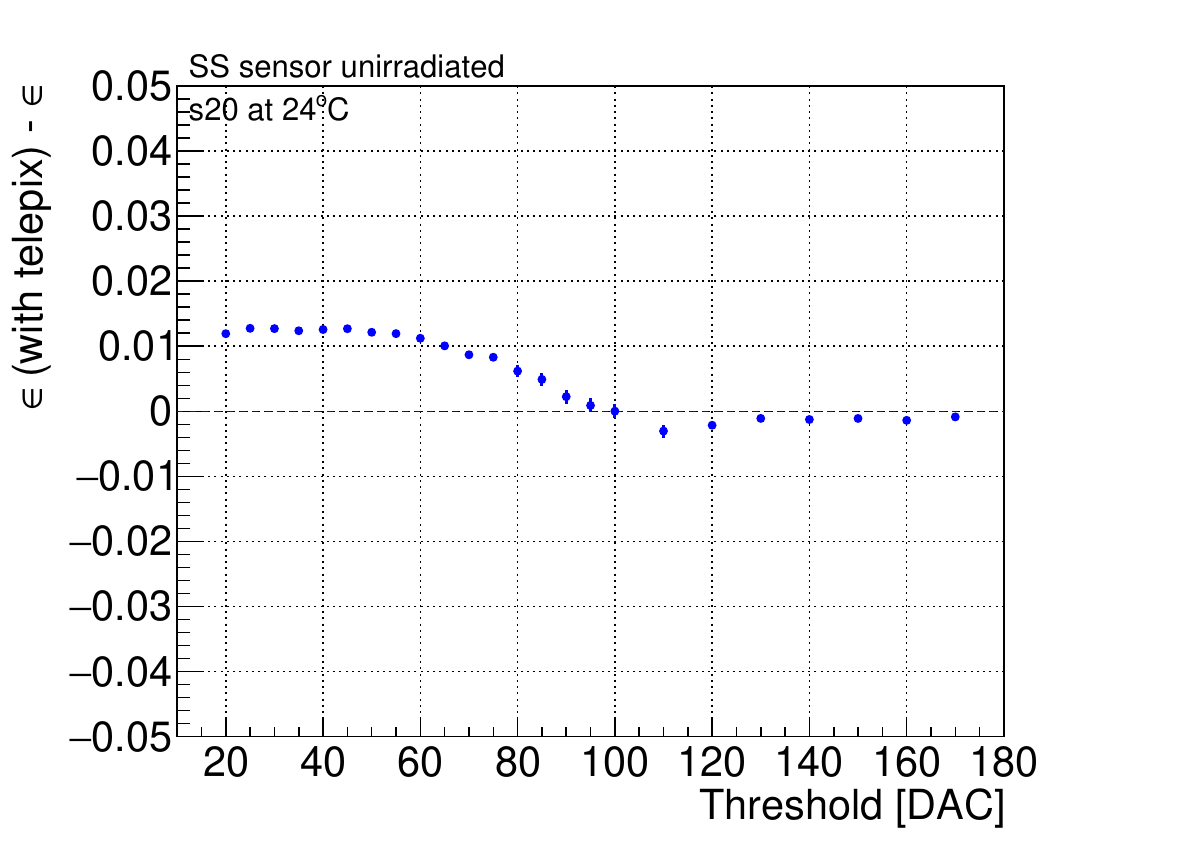}
		\caption{Difference of measured efficiency with and without \tp.}
	\end{subfigure}	
\caption{Comparison of measured efficiency with and without \tp.}
\label{fig:usITkStrip}
\end{figure}

The ATLAS inner tracker (ITk) \cite{itk} is a full silicon detector upgrade of the ATLAS tracking system for the high luminosity phase of the Large Hadron Collider (HL-LHC).
To keep the tracking performance in the expected harsh environment and balance the cost, micro-strip modules are used in the outer barrel and end-cap layers of ITk, operated in a binary readout mode. 
The modules are required to have an operating threshold window with hit detection efficiency above \SI{99}{\percent} at a noise occupancy below 0.1\%. Different types of modules have been tested in the \dtwo facility.
Already \SI{1}{\percent} of fake tracks causes modules to not fulfil this stringent requirement. 
As the modules are significantly larger than the ADENIUM telescope's active area, only the track timing feature of \tp is used in the analysis.
Figure \ref{fig:usITkStrip} shows the comparison of the measured efficiency with and without \tp as timing plane as a function of the detection threshold of an non-irradiated short strip module. 
In the low threshold region, an improvement in the determined efficiency of above \SI{1}{\percent} is observed, resulting in a significant operating region, that would be hidden without the timing capabilities of \tp. 

\subsection*{RD50-MPW4 - region of interest trigger and timing mode}

The \emph{RD50-MPW4} \cite{PILSL2024} is an HV-CMOS pixel sensor fabricated in a \SI{150}{nm} process by \emph{LFoundry}. It consists of a \SI[parse-numbers=false]{64\times64}{pixel} matrix with a pitch of \SI[parse-numbers=false]{62\times62}{\micro m^2}, resulting in an active area of $\approx$~\SI[parse-numbers=false]{4\times4} {mm^2}. 
The pixels are equipped with a time-stamping logic, which allows for \SI{25}{ns} time binning.
The performance was evaluated in a test beam at the \dtwo facility utilizing the ADENIUM telescope and the AIDA-TLU. 
As the ALPIDE sensors are substantially larger than the RD50-MPW4, an ROI trigger as input for the TLU over the ''standard'' operation employing scintillators was favourable, as otherwise, most tracks would not intersect the DUT and therefore occupy unnecessary storage and slow down the data analysis. 
Furthermore, the timing performance of the sensor should be assessed.
\tp is suited well for both by providing a region of interest trigger and allowing for track timestamping during offline analysis.

\begin{figure}
\centering
	\begin{subfigure}{0.32\columnwidth}
		\includegraphics[width=\columnwidth]{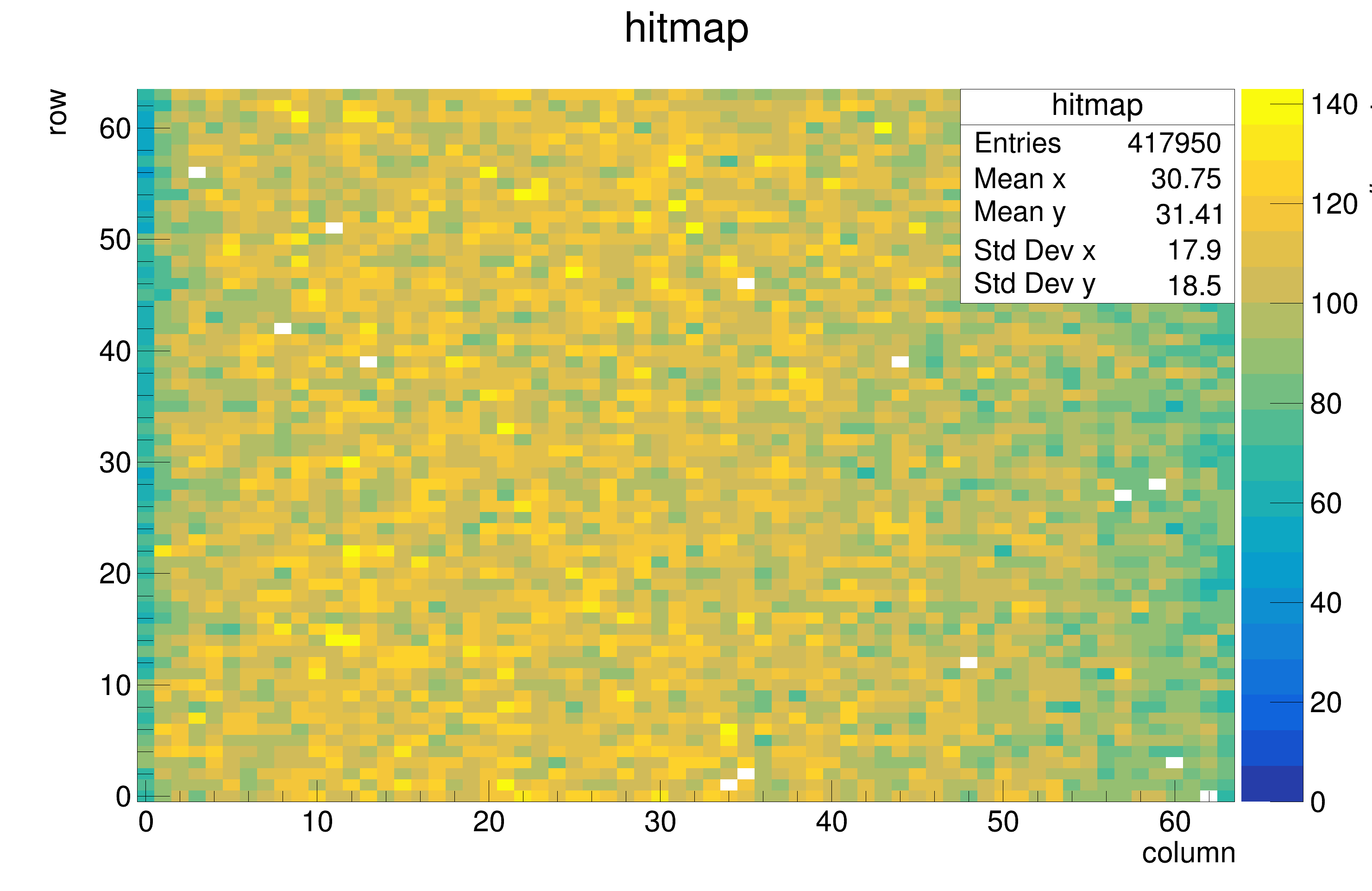}
		\caption{Hitmap of the \emph{RD50-MPW4}.}
	\end{subfigure}	
	\begin{subfigure}{0.32\columnwidth}
		\includegraphics[width=\columnwidth]{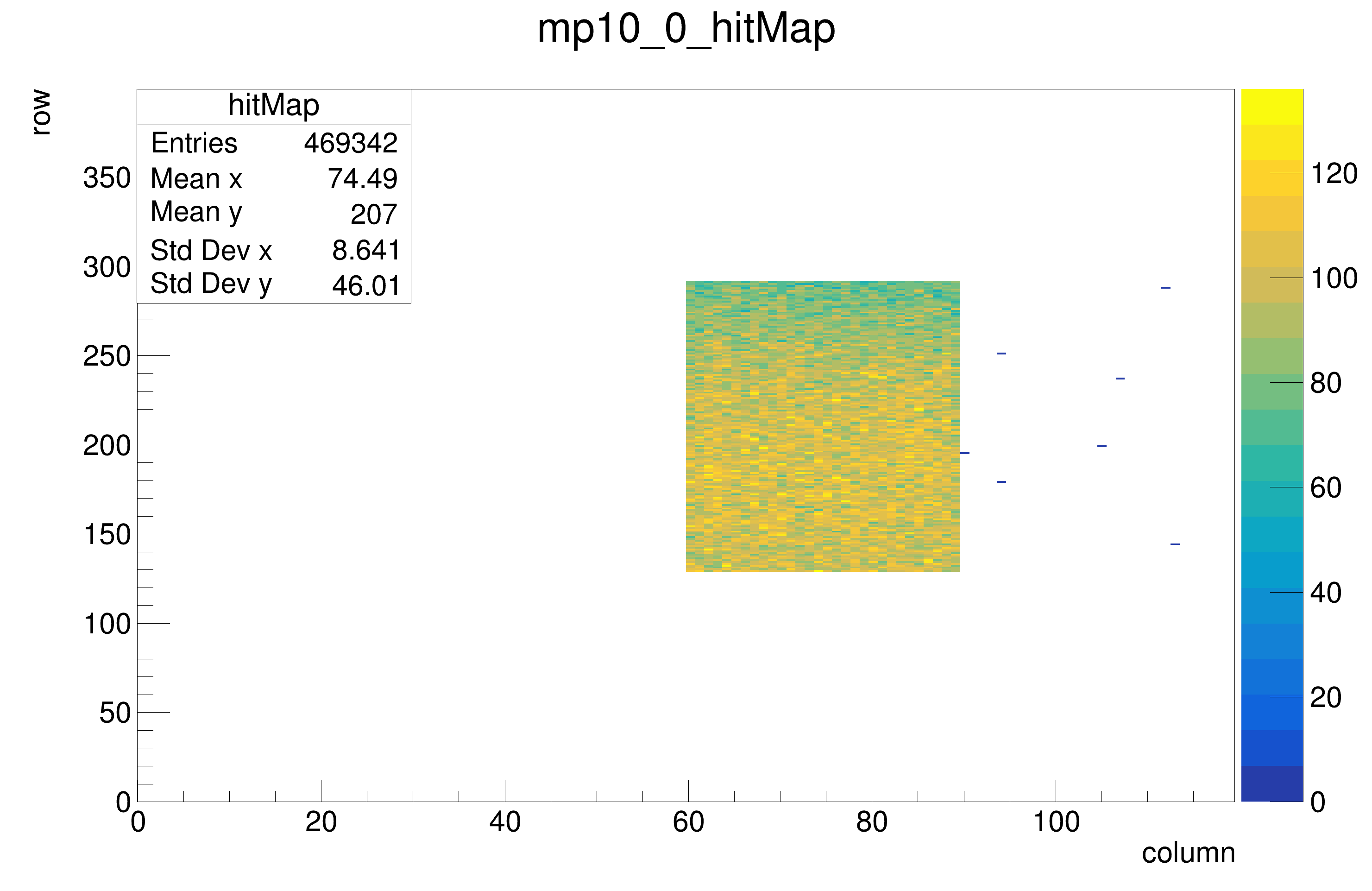}
		\caption{Hitmap of \tp with active ROI.}
	\end{subfigure}	
	\begin{subfigure}{0.32\columnwidth}
		\includegraphics[width=\columnwidth]{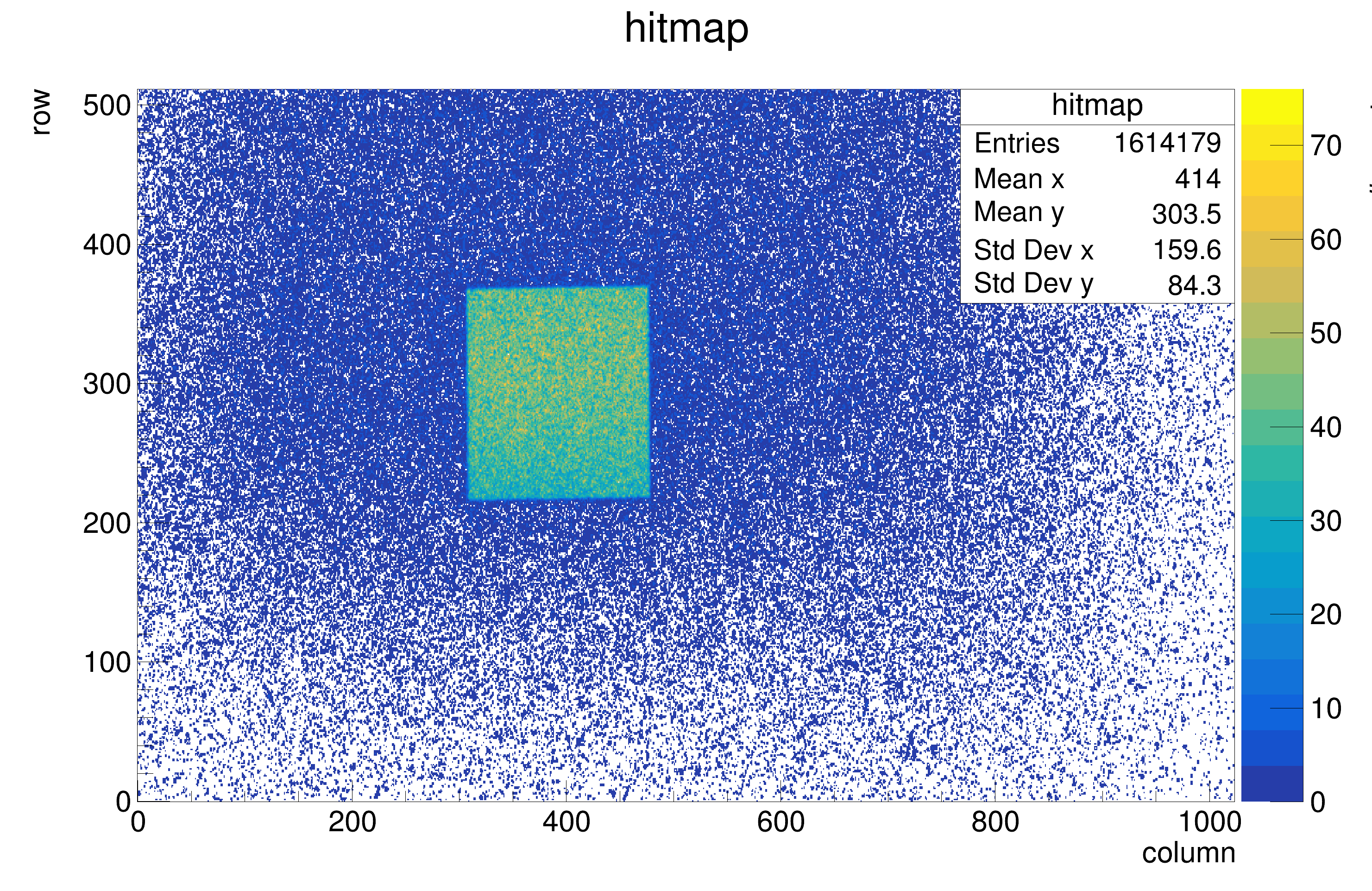}
		\caption{Hitmap of the most downstream telescope plane.}
	\end{subfigure}
\caption{Hitmaps of various sensors in the setup showcasing the effect of the ROI trigger operation of \tp.}
\label{fig:usHitmaps}
\end{figure}

\begin{figure}
\centering
	\begin{subfigure}{0.49\columnwidth}
	\includegraphics[width=\columnwidth]{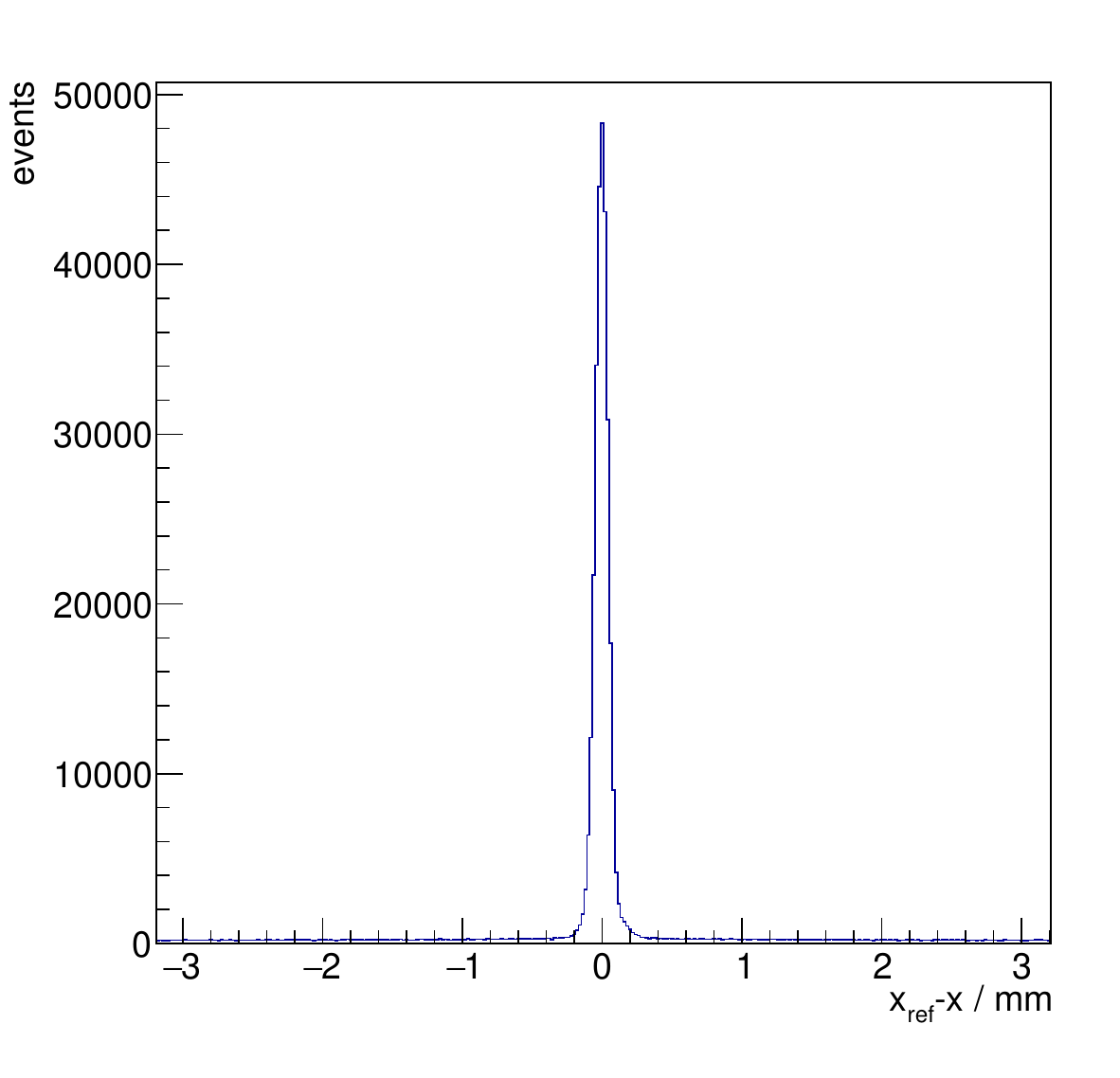}
	\caption{Spatial correlations in X of the DUT with the telescope reference plane and ROI triggering}
\label{fig:usCorrX}
	\end{subfigure}	
		\begin{subfigure}{0.49\columnwidth}
	\includegraphics[width=\columnwidth]{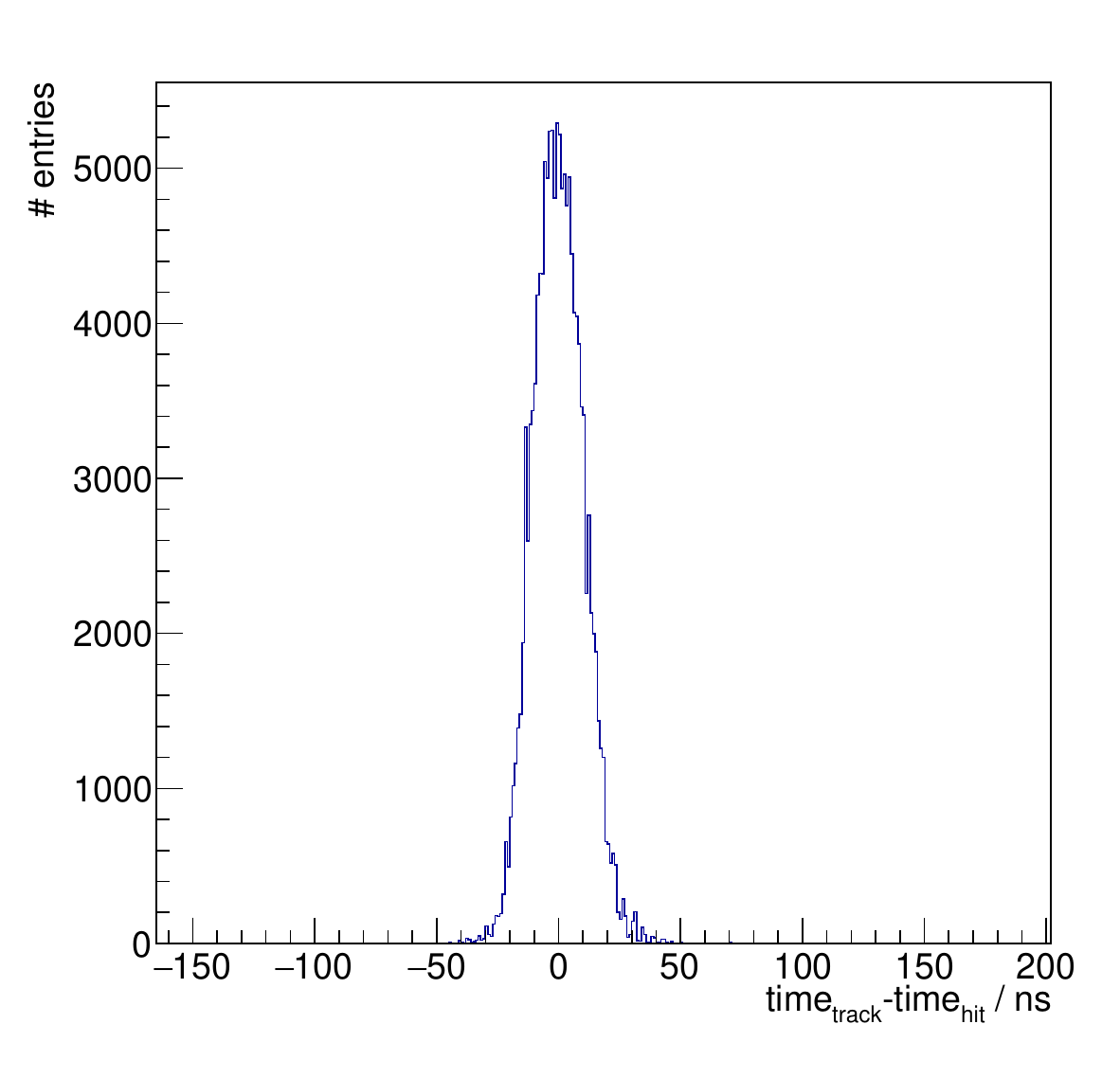}
	\caption{Difference between the hit-time at the DUT and the time of the track, which is assigned from the \tp timestamp.}
	\label{fig:usTimeRes}
	\end{subfigure}	
	\caption{Exemplary test beam results of the RD50-MPW4 highlighting the benefits of employing the \tp as an additional telescope plane.}
	\label{fig:usMpw4Results}
\end{figure}

All pixels of \tp that did not overlap with the DUT were masked and the \hb enabled for all others. 
The effect of the ROI trigger operation is clearly visible in the hitmaps in figure~\ref{fig:usHitmaps}. 
The correlations between the hit position at the DUT and the one at the telescope plane, adjacent to the DUT, in figure~\ref{fig:usCorrX} show minimal background.
\tp can be used during analysis in track reconstruction to further suppress ambiguities and increase the on-DUT tracking efficiency.
Requiring a hit on \tp for a track to be accepted, \SI{80}{\percent} of all tracks intersect the DUT, compared to \SI{63}{\percent} indicating, that there are still events with additional particles that did not pass the trigger window of \tp, that is slightly larger than the DUT allowing for high statistics also at the edges of the DUT.
 \tp efficiently suppresses these unwanted tracks.
Assigning the timestamp of the \tp hit data to the tracks and comparing it to the DUT time stamp allows to depict the temporal resolution of the DUT, by evaluating the time residuals, see figure~\ref{fig:usTimeRes}. The observed value is slightly higher than the basic estimation of $\SI{25}{ns}/ \sqrt{12} \approx \SI{7.2}{ns}$. This is currently under investigation profiting largely from the \tp track time stamping.

\section{Conclusion}
\tp shows excellent performance with efficiencies above \SI{99}{\percent} over a large threshold range and a time resolution of \SI{3.844\pm0.002}{ns} without applying any threshold trimming or offline corrections. 
The fast hit-or which can be configured to any pixel arrangement has a time resolution of \SI{2.28\pm 0.01}{ns} with an absolute delay of \SI{15.32\pm 0.01}{ns} compared to a trigger scintillator.
It is fully integrated with EUDAQ2 and the AIDA-TLU. 
The system is successfully provided to first users serving as timing and region of interest trigger layer. Including \tp in the data taking and analysis improves their results and data taking efficiency. 
By overcoming previous limitations of the telescopes at the DESY II test beam facility, \tp will be key to support future test beam characterizations of novel sensors. \\
Offline time corrections as well as systematic optimization of chip settings to maximize the performance are ongoing and are likely to further improve the presented results.
In parallel, the radiation tolerance is studied.
\section*{Acknowledgments}
This project has received funding from the European Union’s Horizon 2020 Research and Innovation programme under GA no 101004761. 
We acknowledge support by the Deutsche Forschungsgemeinschaft (DFG, German Research Foundation) under Germany’s Excellence Strategy – EXC 2121 ‘‘Quantum Universe’’ – 390833306.
The measurements leading to these results have been performed at the Test Beam Facility at DESY Hamburg (Germany), a member of the Helmholtz Association (HGF).

\bibliography{biblio}

\begin{thebibliography}{10}

\bibitem{DIENER2019265}
R.~Diener et. al.
\newblock {The DESY II test beam facility}.
\newblock {\em NIM-A}, 922:265--286, 2019.

\bibitem{jansen2016}
H.~Jansen et~al.
\newblock Performance of the {EUDET}-type beam telescopes.
\newblock {\em Eur. Phys. J. Tech. Instr.}, 3(1):7, 10 2016.

\bibitem{huguo2010}
C.~Hu-Guo et~al.
\newblock First reticule size {MAPS} with digital output and integrated zero
  suppression for the {EUDET-JRA1} beam telescope.
\newblock {\em Nucl. Instrum. Methods Phys. Rev. A}, 623(1):480--482, 2010.
\newblock 1st International Conference on Technology and Instrumentation in
  Particle Physics.

\bibitem{baudot2009}
J.~Baudot et~al.
\newblock {First test results of MIMOSA-26, a fast CMOS sensor with integrated
  zero suppression and digitized output}.
\newblock In {\em Nuclear Science Symposium Conference Record (NSS/MIC), 2009
  IEEE}, pages 1169--1173, Oct 2009.

\bibitem{Liu:2019wim}
Y.~Liu et~al.
\newblock {{EUDAQ2\textemdash{}A flexible data acquisition software framework
  for common test beams}}.
\newblock {\em JINST}, 14(10):P10033, 2019.

\bibitem{Baesso_2019}
P.~Baesso, D.~Cussans, and J.~Goldstein.
\newblock {The {AIDA}-2020 {TLU}: a flexible trigger logic unit for test beam
  facilities}.
\newblock {\em Journal of Instrumentation}, 14(09):P09019--P09019, sep 2019.

\bibitem{Obermann_2014}
T~Obermann, M~Backhaus, F~Hügging, H~Krüger, F~Lütticke, C~Marinas, and
  N~Wermes.
\newblock Implementation of a configurable fe-i4 trigger plane for the aida
  telescope.
\newblock {\em Journal of Instrumentation}, 9(03):C03035, mar 2014.

\bibitem{AUGUSTIN2023167947}
{Heiko Augustin and Sebastian Dittmeier and Jan Hammerich and Adrian Herkert
  and Lennart Huth and David Immig and Ivan Perić and André Schöning and
  Adriana Simancas and Marcel Stanitzki and Benjamin Weinläder}.
\newblock Telepix – a fast region of interest trigger and timing layer for
  the eudet telescopes.
\newblock {\em Nuclear Instruments and Methods in Physics Research Section A:
  Accelerators, Spectrometers, Detectors and Associated Equipment},
  1048:167947, 2023.

\bibitem{PERIC2007876}
Ivan Perić.
\newblock A novel monolithic pixelated particle detector implemented in
  high-voltage cmos technology.
\newblock {\em Nucl. Instrum. Meth. A}, 582(3):876--885, 2007.
\newblock VERTEX 2006.

\bibitem{SCHIMASSEK2021164812}
R.~Schimassek et~al.
\newblock {Test results of ATLASPIX3 — A reticle size HVCMOS pixel sensor
  designed for construction of multi chip modules}.
\newblock {\em Nucl. Instrum. Meth. A}, 986:164812, 2021.

\bibitem{TWEPP2017}
H.~Augustin et~al.
\newblock {The MuPix Telescope: A Thin, High-Rate Tracking Telescope}.
\newblock {\em Journal of Instrumentation}, 12(01):C01087, 2017.

\bibitem{Dittmeier2018}
S.~Dittmeier.
\newblock {\em {Fast data acquisition for silicon tracking detectors at high
  rates}}.
\newblock PhD thesis, Heidelberg University, 2018.

\bibitem{Huth2018}
L.~Huth.
\newblock {\em {A High Rate Testbeam Data Acquisition System and
  Characterization of High Voltage Monolithic Active Pixel Sensors}}.
\newblock PhD thesis, Heidelberg University, 2018.

\bibitem{Liu_2023}
Yi~Liu, Changqing Feng, Ingrid-Maria Gregor, Adrian Herkert, Lennart Huth,
  Marcel Stanitzki, Yao Teng, and Chenfei Yang.
\newblock Adenium — a demonstrator for a next-generation beam telescope at
  desy.
\newblock {\em Journal of Instrumentation}, 18(06):P06025, June 2023.

\bibitem{Dannheim_2021}
D.~Dannheim, K.~Dort, L.~Huth, D.~Hynds, I.~Kremastiotis, J.~Kröger,
  M.~Munker, F.~Pitters, P.~Schütze, S.~Spannagel, T.~Vanat, and M.~Williams.
\newblock Corryvreckan: a modular 4d track reconstruction and analysis software
  for test beam data.
\newblock {\em Journal of Instrumentation}, 16(03):P03008, mar 2021.

\bibitem{itk}
{Technical Design Report for the ATLAS Inner Tracker Pixel Detector}.
\newblock Technical report, CERN, Geneva, 2017.

\bibitem{PILSL2024}
B.~Pilsl, T.~Bergauer, R.~Casanova, H.~Handerkas, C.~Irmler, U.~Kraemer,
  R.~Marco-Hernandez, J.~{Mazorra de Cos}, F.R. Palomo, S.~Powell, P.~Sieberer,
  J.~Sonneveld, H.~Steininger, E.~Vilella, B.~Wade, C.~Zhang, and S.~Zhang.
\newblock Characterization of the rd50-mpw4 hv-cmos pixel sensor.
\newblock {\em Nuclear Instruments and Methods in Physics Research Section A:
  Accelerators, Spectrometers, Detectors and Associated Equipment},
  1069:169839, 2024.

\end{thebibliography}

\end{document}